\documentclass[times]{speauth}
\usepackage[latin1]{inputenc}
\usepackage{url}
\usepackage{hyperref} 
\usepackage{breakurl} 
\usepackage{amsmath}
\usepackage{alltt}
\usepackage[outdir=./]{epstopdf}
\usepackage{color}

\newcommand{\cpp}{{\sc{}C}{\tt{}++}}

\begin{document}
\runningheads{D. Colnet and B. Sonntag}{Exploiting array manipulation habits}

\title{Exploiting array manipulation habits to optimize garbage collection and type flow analysis}

\author{Dominique Colnet\affil{1} and Benoît Sonntag\affil{2}}
\address{\affilnum{1}\corrauth\ Université de Lorraine, LORIA, Villers-Lès-Nancy, F-54600, France\break
         \affilnum{2}Université de Strasbourg, LSIIT, Illkirch, F-67412, France}
\corraddr{Dominique Colnet, LORIA, Campus Scientifique, BP 239, 54506 Vand\oe{}uvre-l\`{e}s-Nancy Cedex, France. E-mail: Dominique.Colnet@loria.fr}

\begin{abstract}
A widespread practice to implement a flexible array is
to consider the storage area into two parts: the used area which is 
already available for read/write operations and, the supply area, which is
used in case of enlargement of the array.
The main purpose of the supply area is to avoid as much as possible the reallocation
of the whole storage area in case of enlargement.
As the supply area is not used by the application, the main idea of the paper is to
convey the information to the garbage collector, making it possible to avoid
completely the marking of the supply area.

We also present a simple method to analyze the types of objects
which are stored in an array as well as the possible presence of {\sc{}null} values
within the array.
This allows us to better specialize the work of the garbage collector when
marking the used area, and also, by transitivity, to improve overall results
for type analysis of all expressions of the source code.

After introducing several abstract data types which represent the main arrays concerned
by our technique (i.e. zero or variable indexing, circular arrays and hash maps), we
measure its impact during the bootstrap of two compilers whose libraries are equipped
with these abstract data types.
We then measure, on various software products we have not written, the 
frequency of certain habits of manipulation of arrays, to assess the validity of
our approach.
\end{abstract}


\maketitle



\section{Introduction}
A well known technique for implementing a flexible array is to organize
the storage area into two parts: the memory area currently being used, and
the reserve memory area.
The size of the used area gives the current size of the flexible array, and,
if the program behaves normally, only the used area is accessed
for read / write.
When the array is to be extended, extra space is taken first in the supply area.
If the supply area is exhausted, the whole storage area must then be reallocated in
order to provide for a large enough new storage area, which, most often involves 
copying out the used area.
To limit re-allocations and copies as much as possible, one has to find the right
balance to resize the supply area.
As the supply area is not visible by the user program, zeroing of that area is
not required.
In case of decrease in the size of the flexible array, it is enough to change the
boundary mark between the two areas, leaving all values in the storage unchanged.
Thus, all along the execution, the supply area may grow or shrink, containing whatever
values.

The main idea of this paper is to adapt the garbage collector (GC \cite{jones96a}) so that supply areas,
which correspond to inaccessible objects of the application, be completely ignored
during the marking phase.
Our contributions are summarized below:
\begin{itemize}
\item
Presentation of the abstract data type for flexible arrays indexed from 0.
Type analysis for the content of arrays with detection of the NULL value.
Tuning of the GC to ignore the supply area and to use 
type information.
\item
Adaptation of our technique for other kinds of arrays. 
Abstract data types for user-defined indexing, circular arrays and hash maps.
\item
Evaluation of the garbage collection savings during the bootstrap of the SmartEiffel
compiler, a large program using the previously defined abstract data types.
\item
Measurements of the impact of the type analysis carried out for arrays during the
bootstrap of the Lisaac compiler. Also a large benchmark.
\item
Detailed experimental evaluation of the arrays habits using a suite of programs 
including real-life programs written in C or Java.
\end{itemize}

The results presented here come from the work done 
in the SmartEiffel
project \cite{zendra97a,zendra1999d} but also in the Lisaac
project \cite{sonntag2002a,sonntag2013a}.
SmartEiffel consists of a compiler for the Eiffel language as well as a class
library including all the kinds of arrays described in this article.
The new optimization for marking arrays we present here is to be added to
the previously published results in \cite{colnet98a}.
Lisaac is also a compiler with a large library, for a prototype-based language.
The arrays described in this article are also part of the Lisaac's library.
Both compilers are completely bootstrapped, written with the language they
translate, and both are using the arrays presented here.
The compilation strategy of Lisaac is more advanced than the one of SmartEiffel,
particularly with regard to type analysis (see \cite{sonntag2013a} for details).

The rest of the paper is organized as follows.
In section \ref{sec-flex-base}, we present, using the example of a simple 
flexible array, how it is possible to draw part of the filling up strategy
to optimize both the garbage collection and type analysis.
The foundations being laid, we present in section \ref{sec-other-arrays} how
it is possible to implement similar strategies for other kinds of more complex
arrays.
To show the gain obtained, we present in section \ref{sec-benchmarks}, the
measurements we made during the bootstrap of two compilers, 
SmartEiffel and Lisaac, as well as the measurements for some other large
applications, written with other languages / libraries.
We then discuss in section \ref{sec-discuss-related} the importance of the
choices made by the languages designers to ensure the initialization of
memory areas, and we also present work in connection with this article.
Finally, section \ref{sec-conclusion} concludes this article.

\section{Setting up with Simple Flexible Arrays}\label{sec-flex-base}
\subsection{Abstract Data Type for Flexible Arrays}\label{subsec-flex-adt}
\begin{figure}
\begin{center}
\includegraphics[scale=1.0]{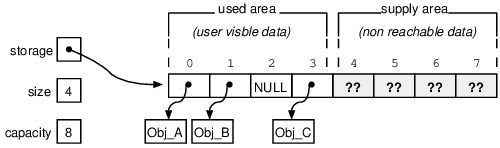}
\caption{Handling a flexible {\sc{}array} with three variables: {\tt{}storage}, 
{\tt{}size} and {\tt{}capacity}. The array is filled up from left to right, 
cell after cell, in order to avoid uninitialized values.}
\label{array-count-capacity}
\end{center}
\end{figure}
Before taking into account other forms of arrays, to start, we now consider the
simple case of a flexible array indexed from zero.
The array can dynamically expand or shrink its size during execution.
The leftmost element is always accessible with index 0.
Figure \ref{array-count-capacity} shows how to manage the array using three variables.
The {\tt{}storage} variable allows access to the elements storage area and,
the {\tt{}capacity} variable stores the allocation size of this area.
The {\tt{}size} variable is the current size of the array as it is seen by the user.
This variable also determines the boundary between the used area 
and the supply area.
The following four operations completely define the abstract data type that
represents the flexible array.
\paragraph{Create(cap)}
The creation operation to be used in order to prepare a new empty 
array with a given {\tt{}capacity} {\tt{}cap}:
{\small\begin{alltt}
  assert (cap >= 0);
  capacity = cap; size = 0; storage = malloc(cap);
\end{alltt}}
\noindent

\paragraph{Extend(obj)}
To extend by one the array on its right, writing {\tt{}obj} in the new slot.
In case of reallocation (i.e. when the supply area is empty), the {\tt{}capacity} is
increased twofold:
{\small\begin{alltt}
  if ( size >= capacity ) \{
     capacity = capacity * 2;
     storage = realloc(storage, capacity);
  \}
  storage[size] = obj;
  size = size + 1;
\end{alltt}}
\noindent

\paragraph{Read(ind)} 
This function returns the object stored at index {\tt{}ind} assuming that the index 
is correct, that is, a valid index in the used area:
{\small\begin{alltt}
  assert ((0 <= ind) && (ind < size));
  return storage[ind];
\end{alltt}}

\paragraph{Write(ind, obj)}
Change the value at index {\tt{}ind} using {\tt{}obj} for the replacement,
assuming that the {\tt{}ind} is a valid index in the used area:
{\small\begin{alltt}
  assert ((0 <= ind) && (ind < size));
  storage[ind] = obj;
\end{alltt}}
The four operations that we have given ({\tt{}Create}
{\tt{}Extend}, {\tt{}Read}, and {\tt{}Write}) 
completely define the abstract data type for flexible arrays.
We will see later how the garbage collector can be made to ignore the supply
area altogether.
Before that, note that these operations force voluntarily the filling up
from left to right by requiring the user to extend the array element
by element.
We made this decision to facilitate the type analysis of the content of
arrays and, in particular, to facilitate the detection of {\sc{}null}.

\subsection{Type Analysis for the Content of Arrays}
Type analysis in object-oriented languages allows a better implementation 
of dynamic dispatch or even, under certain circumstances, the replacement of 
some late binding call sites with direct static calls.
Prediction of the {\sc{}null} values makes it possible to detect statically that
some calls will fail at runtime.
Many research papers have been published regarding type analysis or {\sc{}null} prediction:
\cite{palsberg91a,corney93,dean95a,agesen95a,bacon96a,dean96a,collin97a,zendra97a,fitzgerald2000a,fahndrich2003a,sonntag2013a}.
Previous research papers address type analysis for the {\tt{}self} or {\tt{}this}
variable, instance variables, formal parameters, local and global variables.
When looking for type analysis inside arrays, there is, as far as we know,
nothing published yet.

\begin{figure}
\begin{center}
\includegraphics[scale=1.0]{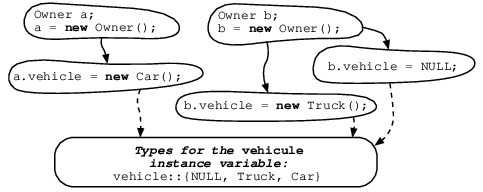}
\caption{Type flow analysis for instance variable {\tt{}vehicle} from class 
{\sc{}owner}. Order of statements is not relevant and the result which is 
\{{\sc{}null}, {\sc{}truck}, {\sc{}car}\}, is for all instances of the class
{\sc{}owner}.}
\label{instance-variables}
\end{center}
\end{figure}

In the Lisaac compiler, we perform a {\it{}type flow analysis} that one must 
not mix up with {\it{data flow analysis}}.
Data flow analysis consists in taking into account the order of statements
to gather a precise piece of information, sometimes even perfect, that is, allowing to
know exactly the reference of the real object or the real value of certain 
expressions.
Often, data flow analysis makes it possible to improve type flow information,
nevertheless, such an accurate type of information is not required to implement our type 
analysis for the content of arrays. 
What we call type flow analysis is the computation, for a given expression, 
of its set of possible dynamic types, {\it{}not taking into account the order 
of statements}.
For instance, using all the reachable assignments in a variable, we compute
all the possible types for that variable.
The analysis is flow insensitive, i.e. the order of reachable assignments does
not matter.
We consider on the whole the set of reachable assignments.

Concerning instance variables, we do not discriminate between the different 
instances from the same class (see figure \ref{instance-variables});
the type information is identical for all instances of a given class.
For a formal method argument, we use the set of reachable effective 
arguments, that is, all reachable calls.
Classically, to obtain the set of dynamic types of a given method call, we
consider all the methods that could be dynamically linked with the receiver 
type set, that is, a kind of dynamic dispatch simulation.
To complete the coverage of all kinds of expressions, we propose in the following
a simple technique to propagate type flow analysis for array read/write operations. 

For the type analysis, all cells of the array are considered as a whole, 
regardless of index variation.
We are using a unique type set for all elements of the array.
As an example, if a {\sc{}truck} is written at index 1, we consider 
that all read accesses, whatever the index used, may yield a {\sc{}truck}.
If another write sequence of a {\sc{}car} exists somewhere in the reachable
code all cells are considered as potential holders of objects 
of type {\sc{}truck} or {\sc{}car}.
More generally, assuming that $S_{array}$ is the type set of the array, as soon as 
one {\it{}index} is possibly assigned with an expression having $S_{expr}$ as 
type set, $S_{array}$ is changed as $S_{array}$ $\cup$ $S_{expr}$.
This is a direct application of the {\it{}array smashing} technique of 
\cite{blanchet2003a} using dynamic types instead of scalar values.

In the rest of the paper, the {\sc{}null} value which represents the lack of
reference is simply considered as a possible type and its occurrence inside 
the type set indicates that the corresponding expression can have the 
{\sc{}null} value.
For instance, in figure \ref{instance-variables}, the set of possible types
of the {\tt{}vehicle} instance variable includes {\sc{}null}.
In the case of flexible array, the choice to start by creating an empty array
which extends cell by cell, facilitates the detection of the {\sc{}null} value.
Indeed, starting from an empty array, if no write operation (i.e. {\tt{}Extend}
or {\tt{}Write}) introduces {\sc{}null}, we statically know that this array
never contains {\sc{}null} at runtime.
With this knowledge, it is possible to customize the marking procedure in order
to focus on the sole used area, completely ignoring the supply area.
Furthermore, inside the used area, it is possible to specialize the marking code
according to type analysis results.
Obviously, this is also true for all method calls applied to objects read from 
the array: {\tt{}Read(ind).methodCall()}.

For the compiler implementation, storing the type flow information for an array
can be done in two ways.
The first keeps all the possible types for an array by making successive unions
with the set of types that an expression inserted in the array can take.
This is a straightforward application of the previous definition.
However, if the set of types of the expression inserted in the array is
incomplete or not known, which can happen at the beginning of the compilation,
then there is another way to proceed.
This second solution involves a list of expressions. 
Every expression ever written to the array is memorized.
This is just a matter of keeping the list of calls to the {\tt{}Extend} and
{\tt{}Write} for a particular array.
The type set of an array can then be computed on demand by propagating the call
to the expressions stored earlier, when all of those expressions get their
set of possible types.

\begin{figure}
\begin{center}
\includegraphics[scale=1.0]{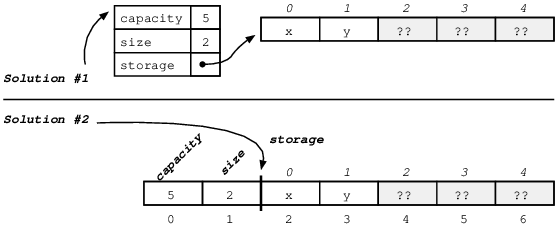}
\caption{Two solutions to wrap all three array manipulation variables,
{\tt{}capacity}, {\tt{}size} and {\tt{}storage}, in order to handle the array
with a single reference.}
\label{impl-array}
\end{center}
\end{figure}

Figure \ref{impl-array} presents two solutions to reference as a whole all the
components of the flexible array.
Solution \#1 is well-adapted to object-oriented programming as it
encapsulates variables {\tt{}capacity}, {\tt{}size} and {\tt{}storage} in 
an object.
This is also very convenient in case of re-allocation of the {\tt{}storage} area
because the reference to the object does not need to be changed.
Solution \#2 avoids the extra indirection to access the {\tt{}storage} area.
The {\tt{}size} and {\tt{}capacity} are to be reached by moving the
pointer by one or two words backward.
Solution \#2 is actually the usual C {\tt{}malloc} solution to memorize the
size of the memory chunk.
That size is necessary in case of a {\tt{}free} call.

\subsection{Avoiding Supply Area Traversal During Garbage Collection}\label{sub-gc-supply-in-arrays}
The most straightforward optimization is to ensure that the garbage collector
scans only the useful area of the array (i.e. the indexes {\tt{}0} to
{\tt{}size - 1}), completely ignoring the supply area.
In the SmartEiffel library, zero indexed flexible arrays are represented with
the {\sc{}fast\_array} class, using solution \#1 of figure \ref{impl-array}.
The compiler allows to partially redefine the marking procedure for the 
{\tt{}storage} area, making it possible to traverse only the useful part,
i.e. the user-visible part of the array.
Thus, the {\sc{}fast\_array} class is equipped with a template method definition
to drive the {\tt{}storage} area scanning:
\paragraph{GarbageCollectorMark()}
{\small\begin{alltt}
  int idx = size - 1;
  while ( idx > 0 ) \{
     MarkOneItemOf(storage, idx); // {\it{}Generated by the compiler (see section 2.4).}
     idx --;
  \}
\end{alltt}}
\noindent

What is done in the SmartEiffel library to ignore the supply area is easily applicable in Java whatever
the category of the GC (i.e. mark and sweep, copying collector or reference counting).
Indeed, Java arrays are managed by the virtual machine and are always initialized to zero.
Arrays of references are already provided with a header that indicates the size of the array and other
information such as those concerning for example the type of the elements.
All write operations in arrays of references go through the execution of the {\tt{}aastore} bytecode.
It is particularly easy to change the Java virtual machine to avoid, during a GC cycle,
the unnecessary traversals of supply areas.
One extra index in each array descriptor is enough for the GC to be informed 
of the boundary of the supply area.
Thus, the {\tt{}aastore} bytecode can be modified as follows:
{\small\begin{alltt}
  aastore(array, value, index) \{
     // {\it{}Unmodified original} aastore {\it{}code here (i.e.} array {\it{}bounds check followed by}
     // {\it{}memory write as well as dynamic type check of the} value{\it{}).}
     // {\it{}As an exception is raised when bounds check fails, the} index {\it{}is always valid in the}
     // {\it{}following extra code:}
     if ( index > gc\_boundary ) \{
        gc\_boundary = index; // {\it{}Supply area shrinked.}
     \}
  \}
\end{alltt}}
The index {\tt{}gc\_boundary} must be initialized to 0 when creating the array.
In Java, for security reasons, all arrays of references are initialized with {\tt{}null}.
It is therefore possible to move the boundary of the supply area
(i.e. {\tt{}gc\_boundary}) without special precautions.
Note that no change is needed for read operations that could occur beyond the index {\tt{}gc\_boundary}.
All values beyond the {\tt{}gc\_boundary} index return {\tt{}null} and do not affect the GC.
Of course, one must also modify the code of the GC so that it takes into account the new 
frontier (i.e. the GC uses {\tt{}array.gc\_boundary} instead of {\tt{}array.length - 1}).

\subsection{Using Type Information to Optimize Garbage Collection of Arrays}\label{sub-gc-type-in-arrays}
The type flow information of the array elements is integrated to the garbage collector as it
is already the case for the objects' attributes.
SmartEiffel generates a specialized and precise marking
function for every object type and without interpretation during execution
(for details about SmartEiffel's GC, see \cite{colnet98a}).
The GC code is generated differently for every application; it is adapted to 
every manipulated object and is integrated in the final executable. 
The marking code has a static (hard-coded) knowledge of the object's structure. 
As an example, an instance variable which is of type integer, character or floating
point number is simply ignored (i.e. no marking code is generated for them), since such
a variable is not a possible path toward another object. 
Only attributes that are references are inspected.
Yet another improvement, thanks to type flow analysis, in case an attribute is 
monomorphic (i.e. when the type set of that attribute has only one type), the 
function call to the marking code is direct.
Polymorphic attributes still need the usual dispatching code to the corresponding
marking code.

This specialization used for attributes (described in \cite{colnet98a}) is now 
also applied to arrays when iterated in their initialized area.
In the particular case of an array that cannot contain {\sc{}null} references,
the {\sc{}null} test is omitted while scanning the array.
It is also possible to avoid a dynamic call if all the elements of an
array have only one possible dynamic type, and if it is the same for all
the elements.
For the general case, type analysis results are used to produce the dispatch
marking code, as we do for ordinary late binding calls.
The internal {\tt{}MarkOneItemOf} call in the loop is processed by the compiler to profit from 
type analysis results as follows:
 
\paragraph{MarkOneItemOf(array, index)}
{\small\tt\begin{tabbing}
~~item = array[index];~~~~~~~~~~~~~~\=// {\it{}Array access into the local} {\tt{}item} {\it{}variable.}\\
\\
~~if ( item == NULL ) \{                  \>// {\it{}Code generated if and only if the null}\\
~~~~~return;                              \>// {\it{}type is in the type set of the array.}\\
~~\}\\
\\
~~switch ( item->dynamic\_type ) \{       \>// {\it{}Switch generated only when the type set}\\
                                          \>// {\it{}has more than one non null element.}\\
\\
~~~~~case TRUCK:                          \>// {\it{}Case branch generated only when the} \\
~~~~~~~~MarkTRUCK(item);                  \>// {\it{}array type set has} {\tt{}TRUCK} {\it{}as element.}\\
~~~~~~~~break;\\
\\
~~~~~case CAR:                            \>// {\it{}Case branch generated only when the} \\
~~~~~~~~MarkCAR(item);                    \>// {\it{}array type set has} {\tt{}CAR} {\it{}as element.}\\
~~~~~~~~break;\\
\\
~~~~~...\\
~~\}\\
\end{tabbing}}

\subsection{Extra Builtins Operations for Flexible arrays}
As might be expected, the {\sc{}fast\_array} class is equipped with many other methods.
Most of them are simple combinations of previously defined methods
of section \ref{subsec-flex-adt}, as such, they require no special support
by the compiler, type analysis results being carried out by
transitivity.
Given our type analysis technique, methods that reduce the size of the visible area
do not need particular compiler support.
As an example, the method to remove the rightmost element can be simply
defined as follow:
\paragraph{RemoveLast()}
{\small\begin{alltt}
  assert (size > 0);
  size--;
\end{alltt}}
\noindent
Note that it is not necessary to reset with {\sc{}null} the cell that has been assigned to the
supply area, as such, that cell will be simply ignored when marking.
Thus, in theory, only the methods defined in section \ref{subsec-flex-adt} are subject
to compiler support for type analysis.
In practice, as it is not always practical to build an array, little by little, cell by cell,
the library features some other builtins.
The {\tt{}Calloc} built-in allows to allocate and initialize an array with just one call.
While the real implementation relies on the efficient C {\tt{}calloc}, the effect,
in terms of type analysis is equivalent to the following code:
\paragraph{Calloc(siz)}
{\small\begin{alltt}
  assert (siz >= 0);
  Create(siz);
  while ( size < siz ) \{ 
     Extend(NULL);
  \}
\end{alltt}}
\noindent
As a consequence, such an allocated new flexible array is a potential holder of
the {\sc{}null} value. 
We needed this built-in to optimize the implementation for all hashing
collections (hash-table, dictionary, hashed set, \ldots).
Implementing a hash table requires creating an array initialized with
{\sc{}null} and having its size match its capacity.
Notice that for performance reasons, {\tt{}calloc} actually uses the {\tt{}memset}
function from the C language, making it possible for the code to use the
specific processor instructions that initialize entire areas in memory.
Also for performance reasons, the {\sc{}fast\_array} class is also equipped
with another built-in to extend arrays with a large number of cells.
That built-in uses the C {\tt{}realloc} function to minimize and optimize data 
moving in memory.

\section{Handling other kinds of arrays}\label{sec-other-arrays}
Zero indexed array (e.g. SmartEiffel's {\sc{}fast\_array}) is not the only one class to 
be handled with a progressive filling up strategy and possibly equipped with high-level
garbage collection tuning.
\subsection{The {\sc{}array} Class}\label{sub-array}
An object of the {\sc{}array} class is an array that can have any integer value for
the left-most index.
Positive value, zero, or a negative value as well.
Except for index translation which needs an extra attribute, gradual
left-to-right filling is appropriate.
The implementation is straightforward and very similar to the one of 
{\sc{}fast\_array}.
Obviously, we get the same benefits in terms of garbage collection and
type analysis.
Let us now see how to handle circular arrays.

\subsection{The {\sc{}ring\_array} Class}
\label{sub-ring-array}
\begin{figure}
\begin{center}
\includegraphics[scale=1.0]{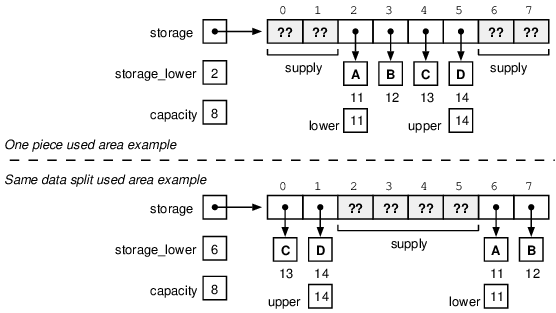}
\caption{Implementation of a {\sc{}ring\_array} with 5 variables. The user's
visible range is {\tt{}[lower}, {\tt{}upper]}. The {\tt{}storage\_lower}
internal variable is the corresponding index of element at {\tt{}lower} index inside 
{\tt{}storage}. Because the used area unfolds circularly starting at any
possible {\tt{}storage\_lower}, there are many possibilities.}
\label{ring-array}
\end{center}
\end{figure}

Adding elements in front of a {\sc{}fast\_array} or in front of an {\sc{}array}
cannot be made efficiently.
As all elements have to be shifted to the right, this makes the operation very
slow.
To address this problem, the SmartEiffel library provides the 
{\sc{}ring\_array} class for managing an array circularly.
In order to keep the flow analysis that can detect the absence of {\sc{}null}s,
it is necessary to adapt the filling up technique.
Figure \ref{ring-array} shows how it is implemented, and the new variables
required.
In the example, the same array with indexes from 11 to 14 is shown twice.
The first shows a case where the used area is in one part and the supply area
is split in two parts.
The second shows the inverse case, but many other situations are possible.

As for previous arrays, the {\tt{}storage} variable points to the beginning of 
the allocated area and the {\tt{}capacity} variable memorizes its size.
The {\tt{}lower} and {\tt{}upper} variables memorize the lower and the upper
bound from the user's point of view, i.e. the publicly visible part of the 
{\sc{}ring\_array}.
The {\tt{}storage\_lower} variable is used internally and matches the
physical index for the {\tt{}lower} indexed element in {\tt{}storage}.
As such, the {\tt{}storage\_lower} variable always stays between 0 and 
{\tt{}capacity - 1}.

The following operations define the abstract data type for the 
{\sc{}ring\_array}, keeping type flow analysis properties.

\paragraph{RingCreate(cap, low)}
Creation of an empty {\sc{}ring\_array}.
The lower index is initialized with {\tt{}low} and the initial
{\tt{}capacity} with {\tt{}cap}:
{\small\begin{alltt}
  assert (cap >= 0);
  capacity = cap; lower = low;
  storage = malloc(cap);
  storage\_lower = 0; upper = low - 1;  
\end{alltt}}
\noindent
As for the creation of non-circular arrays, the initial used area is empty and the
set of possible types is initialized with the empty set.

\paragraph{RingExtend(obj)}
To extend by one the array on its right.
When the supply area is empty, as previously, the {\tt{}capacity} is made twice
as big:
{\small\begin{alltt}
  if ( upper - lower + 1 >= capacity ) \{
     capacity = capacity * 2;
     storage = realloc(storage, capacity);
  \}
  sidx = upper + 1 - lower + storage\_lower;
  if (sidx >= capacity) \{
     sidx = sidx - capacity;
  \}
  storage[sidx] = obj; upper = upper + 1;
\end{alltt}}
\noindent
As for other arrays, as soon as a call of {\tt{}RingExtend} is in the reachable code, all slots 
are considered as possible holders of {\tt{}obj}'s types.

\paragraph{RingPrepend(obj)}
To extend by one the array on its left.
The type flow information is maintained the same way we did for {\tt{}RingExtend}.
Only the index computation in the {\tt{}storage} area is different: 
{\small\begin{alltt}
  if ( upper - lower + 1 >= capacity ) \{
     capacity = capacity * 2;
     storage = realloc(storage, capacity);
  \}
  storage\_lower = storage\_lower - 1;
  if (storage\_lower < 0) \{
     storage\_lower = storage\_lower + capacity;
  \}
  storage[storage\_lower] = obj; lower = lower - 1;
\end{alltt}}

\paragraph{RingRead(ind)}
To access the element which is at index {\tt{}ind}, assuming that {\tt{}ind} is a valid
index in the visible area:
{\small\begin{alltt}
  assert (lower <= ind) && (ind <= upper);
  sidx = ind - lower + storage\_lower;
  if (sidx >= capacity) \{
     sidx = sidx - capacity;
  \}
  return storage[sidx];
\end{alltt}}
\noindent
As this only reads the array and does not modify it, it doesn't change the type
flow information previously gathered.

\paragraph{RingWrite(ind, obj)}
{\small\begin{alltt}
  assert (lower <= ind) && (ind <= upper);
  sidx = ind - lower + storage\_lower;
  if (sidx >= capacity) \{
     sidx = sidx - capacity;
  \}
  storage[sidx] = obj; 
\end{alltt}}
\noindent
As usual, since this adds a new value in the array, the possible types of 
{\tt{}obj} must be all added into the set of possible types contained in the 
array.

And finally, to complete the description of the abstract data type for 
{\sc{}ring\_array}, we have to make the GC aware of the visible 
area with the following template definition:
\paragraph{GarbageCollectorRingMark()}
{\small\begin{alltt}
  int cnt = upper - lower + 1;
  int idx = storage\_lower - 1;
  while ( cnt >= 0 ) \{
     idx ++;
     if (idx >= capacity) \{
        idx = idx - capacity;
     \}
     MarkOneItemOf(storage, idx);
     cnt --;
  \}
\end{alltt}}
\noindent
Again, the {\tt{}MarkOneItemOf} call is generated by the compiler and
customized according to type analysis results.
In the best case, there is only one possible type and the exact marking code
for only one kind of element is inlined.

\subsection{Hash-Tables or Hash-Maps Using an Array}
\label{sub-hash-map}
\begin{figure}
\begin{center}
\includegraphics[scale=1.0]{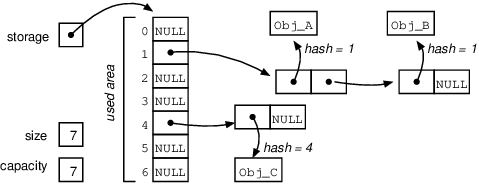}
\caption{Using an array as a {\it{}hash map} to implement a set. All the 
{\tt{}capacity} of the {\tt{}storage} area is used, consequently, the supply
area is empty. {\sc{}null} values inside the array are meaningful.}
\label{hashed_set}
\end{center}
\end{figure}
More complex data structure using hashes, as for example hash sets, are
using a primary array that might contain {\sc{}null} references or a reference
to a storage cell (figure \ref{hashed_set}).
Access to this array uses the {\it{}hash code} of the targeted object.
If the array contains a {\sc{}null} at the index corresponding to the
{\it{}hash code}, it means that the object is not present in the set.
If the array contains a valid reference to a storage cell instead, the object
can then be searched in the linked list composed of cells to account for any
collisions.
In order to create the primary array, the built-in {\tt{}calloc} is used.
As the primary array does not contain a reserved area for future expansion, the
GC normally works by exploring the whole array.
And as it is absolutely common for the primary array to contain {\sc{}null}s, 
the use of {\tt{}calloc} is clearly the best choice as it implies the presence
of {\sc{}null} in the set of possible types.

\section{Benchmarks and Measurements}\label{sec-benchmarks}
\subsection{Garbage Collection for Arrays in SmartEiffel}\label{sub-bench-eiffel}
We equipped the source code of the SmartEiffel's GC with extra code in 
order to examine the impact of arrays on memory footprint.
To have a significant execution we have used the whole source code of the 
SmartEiffel compiler itself, that is 180,000 lines of Eiffel source code during
its own bootstrap.
The self recompilation of the compiler is a very good benchmark since it uses
a lot of arrays and requires about 330 Mb of memory during the process.
Moreover, the GC is triggered 32 times while compiling the 
compiler\footnote{The heap size is automatically tuned with the results of 
the memory collection.}.

For the following measurements, only arrays of references are considered, 
since other arrays, like for example arrays of integers or arrays of characters
are not directly concerned by our type flow analysis technique.
Moreover, the SmartEiffel GC does not even scan the content of 
arrays of scalars.

We modified the marking procedure for the content of arrays so as to count the 
number of marked arrays during one recompilation. 
The measurement shows that the GC processes 6,399,198 arrays.
The total size of the corresponding used area scanned is 12,548,963 
cells.
As the total {\tt{}capacity} of processed arrays is 21,714,957 cells and
since the supply area is not scanned, the GC avoids scanning 9,165,994 cells,
that is, a gain of 42\%.


\begin{figure}
\begin{center}
\includegraphics[scale=1.2]{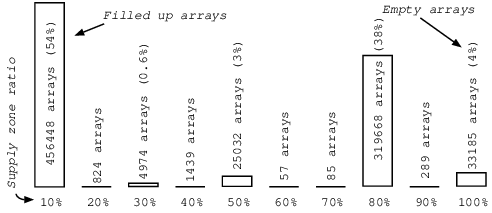}
\caption{
Bootstrap of SmartEiffel. Distribution of arrays according to the relative size of the supply area. Only
arrays of references are considered for figures \ref{gc-supply}, \ref{gc-capacity-ratio} and
\ref{gc-supply-ratio} (same run).}
\label{gc-supply}
\end{center}
\end{figure}

In order to gather more information, we examined the distribution of arrays during
the very last run of the GC, just before the final exit of the program.
During the last run of the GC, among the 838,681 marked arrays, 31,859 are
actually empty arrays.
These empty arrays take direct advantage of our optimization since they are 
marked in a constant time.
By contrast, there are 453,215 arrays which are completely filled up and, 
as a consequence, need to be entirely scanned.
Figure \ref{gc-supply} displays the distribution of arrays according to the 
relative size of the supply area.
The filled up arrays are represented in the left-most bar and the empty 
arrays are in the right-most bar.
In figure \ref{gc-supply}, the largest bar is for arrays having less than
10\% of supply area which seems to indicate that SmartEiffel does not waste 
too much memory using arrays with large supply areas.

\begin{figure}
\begin{center}
\includegraphics[scale=1.2]{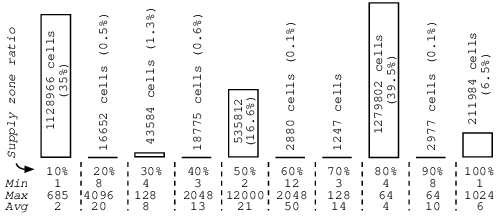}
\caption{Bootstrap of SmartEiffel. Following figure \ref{gc-supply} with the same distribution, the 
relative size of capacities in number of cells.}
\label{gc-capacity-ratio}
\end{center}
\end{figure}

\begin{figure}
\begin{center}
\includegraphics[scale=1.2]{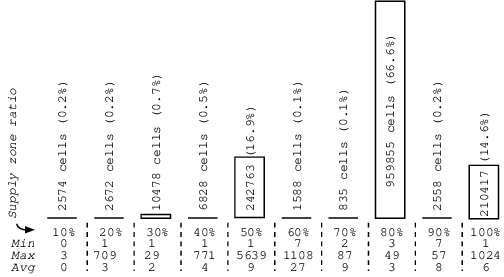}
\caption{Bootstrap of SmartEiffel. Following figure \ref{gc-supply} with the same distribution, the 
relative size of supply areas in number of cells.}
\label{gc-supply-ratio}
\end{center}
\end{figure}

In addition, during the last run of the GC, we also measured that the total 
{\tt{}capacity} of storage was 3,242,679 cells including 1,440,568 cells
in the supply areas.
Using the distribution of arrays from figure \ref{gc-supply}, figure 
\ref{gc-capacity-ratio} gives the corresponding capacities for storage. 
The values {\it{}Min}, {\it{}Max} and {\it{}Avg} give respectively for each
family (the 10\% family, the 20\% family, the 30\% family, etc.), the smallest 
capacity, the largest capacity and the average capacity.
The average size of arrays is small and filled up arrays represent 35\% 
of the total memory used for arrays. 
Still using the same distribution of arrays, figure \ref{gc-supply-ratio} 
shows the corresponding sizes for the supply areas.
The 80\% family represents the largest part of the memory used for the supply
areas, which probably correspond to arrays used as temporary buffers, growing
and shrinking all along the execution.

If one compares the memory used for arrays of references (3,242,679 references) 
with the total memory footprint during bootstrap (330 Mb, that is 
82$\times$10$^{6}$ references), the memory used for arrays of references 
represents only 4\% of that memory.
Considering only the {\tt{}supply} areas, it represents even less, actually 1.7\% 
of the total memory footprint.
Thus, during bootstrap, the total memory used for arrays of references is 
so small compared with the memory used for other objects that the gain of runtime
due to optimized array marking is probably low. 
We modified the GC of SmartEiffel in order to force scanning of the supply 
areas but, as expected, we did not notice any discernible modification of 
runtime, either of memory used or in the number of GC calls.

\subsection{Type Flow Analysis in the Lisaac Compiler}\label{sub-bench-lisaac}
While trying to perform a true global system analysis, the lack of type
analysis concerning the content of arrays has consequences in the rest of the 
analysis.
For instance, as soon as a variable is assigned with a reference read from an 
array, the lack of type information on the content of the array is echoed
on the information that was previously gathered for this variable.
If this variable is then copied into another variable or passed as an argument,
the unclear information about the array content propagates rapidly.
Array read or array write operations are quite frequent and it is thus 
important to collect, then to transmit, type flow information concerning 
array contents.  

The Lisaac compiler carries out a global type flow analysis for 
all kinds of variables: formal parameters, local, global and instance variables.
We also implemented our type flow analysis with {\sc{}null} detection inside
arrays thus making type flow analysis really global.
The gathered information for arrays may impact all kinds of variables and 
transitively, all kinds of expressions, from methods to methods without any 
border.
We have used the source code of the Lisaac compiler itself, 53,000 lines
of Lisaac source code, as a benchmark for the following measurements.

\begin{figure}
\begin{center}
\includegraphics[scale=1.2]{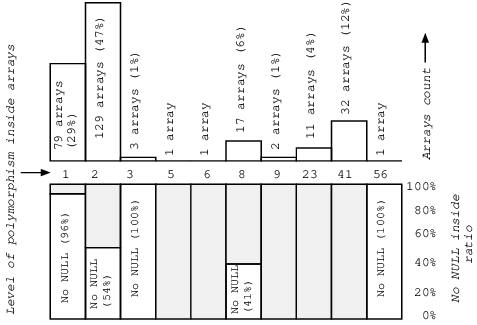}
\caption{Bootstrap of Lisaac. Compile time information regarding types of arrays. The upper part gives
the distribution of arrays according to the number of elements in type sets.
The lower part gives the corresponding ratio of arrays without {\sc{}null}.}
\label{array-type-flow}
\end{center}
\end{figure}
We modified the compiler in order to inspect the types of variables after 
type flow analysis completion.
For the measurement, we kept only variables whose type is an array of 
references.
Arrays of scalar values (arrays of integers, arrays of characters or arrays of 
floats) are not considered because there is no possible polymorphism inside 
such arrays.
Furthermore, only arrays of references may hold {\sc{}null} values.
As explained previously,  {\sc{}null} is considered as a possible element
in the set of types which describe cell contents.
Figure \ref{array-type-flow} shows the distribution of type sets according 
to the number of items in each set.
The level of polymorphism is variable from 1 to 56 for the compiler source code.
As shown in the lower part of the figure, many arrays are statically detected
without possible {\sc{}null} inside cells.
For the source code of the compiler, this stands for 56.2\% of arrays.
Also on figure \ref{array-type-flow}, the left-most bar in the upper part 
indicates that 79 variables are holding arrays with only one kind of 
elements (i.e. there are 79 variables holding {\it{}monomorphic} arrays).
The corresponding bar in the lower part indicates that 96\% of those arrays
never hold {\sc{}null} values.
As an example, for such an array which contains only {\sc{}truck}, the marking function 
generated by the compiler is:
\paragraph{MarkOneItemOf(array, index)}
{\small\tt\begin{tabbing}
~~MarkTRUCK(array[index]);~~~\=// {\it{}Direct static call of the marking function (no NULL check).}\\
\end{tabbing}}

The upper right-hand bar of figure \ref{array-type-flow} shows that only one 
variable gets an array type with 56 kinds of objects mixed together in its 
cells.
A close look at the source code taught us that this variable was actually an 
instance variable.
Thus, at runtime, each object holding that instance variable will have its own
megamorphic array and, as indicated by the lower right-hand bar, all those 
arrays will never include the {\sc{}null} value.

\begin{sidewaysfigure}
\begin{center}
\includegraphics[scale=1.2]{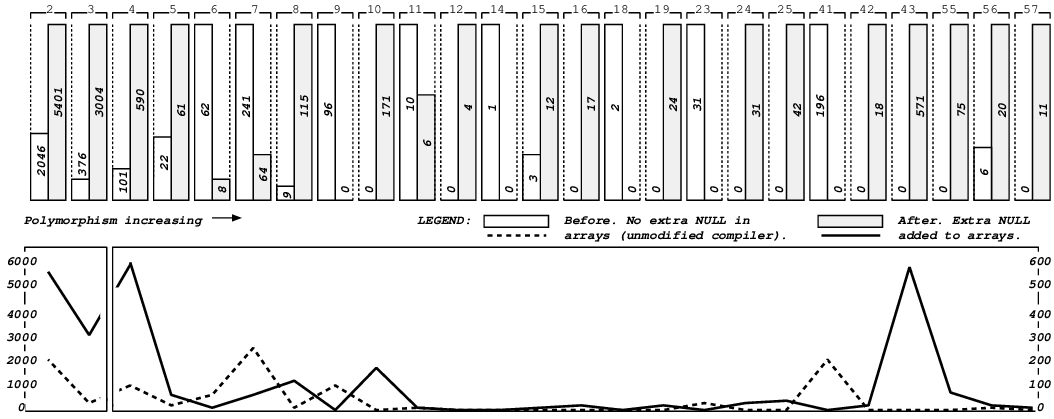}
\caption{Bootstrap of Lisaac. {\it{}Before} and {\it{}After} compiler modification. Adding extra
{\sc{}null} to each type of array to view the propagation of the wrong 
information into local variables type sets. 
The number of local variables is indicated inside each bar (upper part).}
\label{local-type-flowNA}
\end{center}
\end{sidewaysfigure}
\begin{sidewaysfigure}
\begin{center}
\includegraphics[scale=1.2]{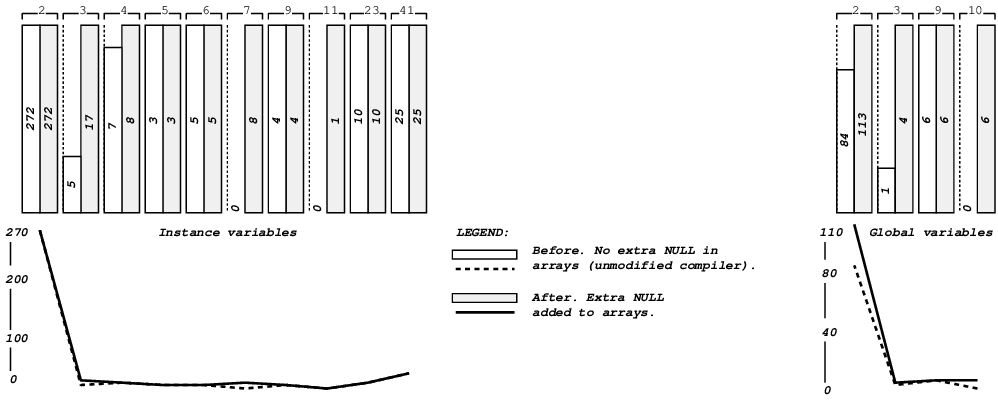}
\caption{Bootstrap of Lisaac. {\it{}Before} and {\it{}After} compiler modification. Adding extra 
{\sc{}null} to each type of array to view the propagation of the wrong 
information into instance variables (left) and global variables (right).}
\label{globattr-type-flowNA}
\end{center}
\end{sidewaysfigure}

To examine the propagation of the type flow information from arrays
to local variables, we modified the compiler in order to add artificially
the {\sc{}null} value inside all sets of types for all arrays.
Among the 10,429 local variables, before the modification of the compiler,
only 3,202 have the {\sc{}null} value inside their type set, that is to say 
30.7\% of the local variables.
Artificial addition of the {\sc{}null} value inside arrays infects 7,043
local variables, making 98.2\% of local variables possible {\sc{}null} holders.
This experimentation shows that the result of type flow information for arrays
directly impacts on the type flow information for local variables.
Figure \ref{local-type-flowNA} gives the detail of the {\sc{}null} propagation,
local variables being sorted according to their level of polymorphism which 
varies from 2 to 57.

We then performed the same measurement for global variables and instance 
variables.
Before the artificial addition of the {\sc{}null} value inside all arrays, 70.5\% 
of global variables were already {\sc{}null} holders.
After the modification, 98.5\% of global variables became {\sc{}null} holders.
For instance variables, the {\sc{}null} propagation changes from 93.2\% of
{\sc{}null} holders to 99.4\%.
Figure \ref{globattr-type-flowNA} details the {\sc{}null} propagation for
global variables and instance variables.

\begin{sidewaysfigure}
\begin{center}
\includegraphics[scale=1.2]{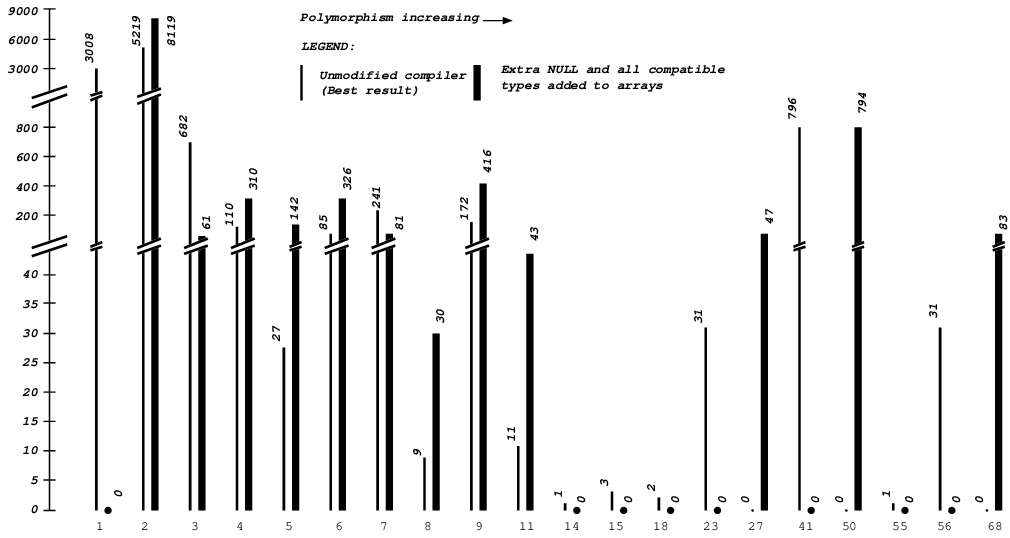}
\caption{Bootstrap of Lisaac. Propagation into local variables of array type sets saturation: 
{\sc{}null} added as well as all compatible types.}
\label{local-type-flowAA}
\end{center}
\end{sidewaysfigure}

To continue the study of array type flow propagation, we then artificially 
added into the type sets of arrays, not only the {\sc{}null} value, but also 
all compatible subtypes of the elements.
For instance, with an array of {\sc{}vehicle}s, all encountered subtypes of 
{\sc{}vehicle} are artificially added into the type set: {\sc{}car} is added, 
{\sc{}truck} is added, {\sc{}bike}, etc., not forgetting to add the {\sc{}null}
value.
For local variables, before the modification, there were 17 different type set
sizes in the whole source code of the compiler.
After the modification, only 12 different sizes remain.
Detailed distribution presented in figure \ref{local-type-flowAA} shows that 
the number of sets decreases and that these sets are bigger, thus indicating 
that the accuracy of type flow analysis for local variable is drastically 
impacted by arrays.

\subsection{Memory Behavior Analysis of Some Other C Programs}\label{sub-bench-other}

\begin{figure}
\begin{center}
\includegraphics[scale=1.0]{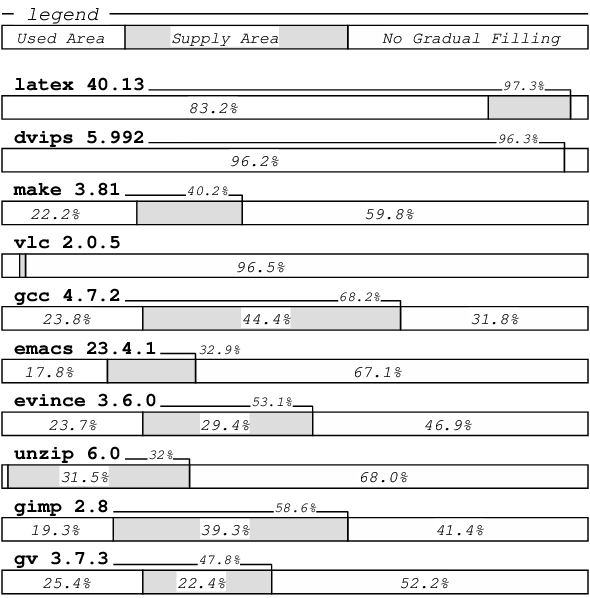}
\caption{Detection of memory blocks which are filled gradually. Relative memory used for all arrays of pointers. Measured just before the program exits.}
\label{gimp-gcc-etc-1}
\end{center}
\end{figure}

\begin{figure}
\begin{center}
\includegraphics[scale=1.0]{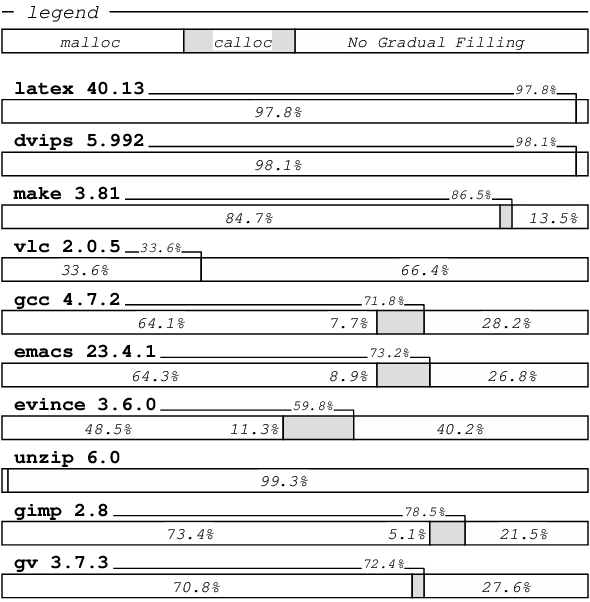}
\caption{Detection of memory blocks which are filled gradually. Relative number of arrays of pointers. Useless {\tt{}calloc} detection. Measured just before the program exits.}
\label{gimp-gcc-etc-2}
\end{center}
\end{figure}

In order to inspect other software products than SmartEiffel and Lisaac,
we analyzed the memory behavior of several other applications
(see figures \ref{gimp-gcc-etc-1} and \ref{gimp-gcc-etc-2}).
To do this, we have developed a new module for Valgrind \cite{bond2007} in
order to observe the allocation of memory blocks that may correspond to arrays
of pointers.
All reads and writes in these blocks are dynamically analyzed
until the end of execution.
Each time a read or write operation involves another data size than
the memory pointer 
size\footnote{The tests were performed on a 64-bit architecture.
We also used the function {\tt{}my\_get\_chunk} from Valgrind which allows to check that
a value corresponds to a pointer to a block in the heap.},
the corresponding block is excluded from the
analysis because we want to study the sole memory devoted to arrays
of pointer, not the memory used for other objects.

The objective is to determine if a memory block is filled gradually or not.
We only detected the gradual left-to-right filling up, from the beginning
of the block.
The other gradual filling strategies, such as for example, the
right-to-left filling up have not been investigated.
Just after the allocation of each new block (i.e. {\tt{}malloc} 
or {\tt{}calloc}), the supply area occupies the whole memory of
that block.
Each new block is initially classified into the gradual-filling category.
A block is kept in the gradual-filling category only if the first write
operation is performed into the leftmost slot of the supply area and so on.
Once a write operation is out of the used area and is not equivalent
to the {\tt{}Extend} operation of section \ref{subsec-flex-adt},
the corresponding block is definitively removed from the gradual-filling
category.

Figure \ref{gimp-gcc-etc-1} shows the results in terms of memory space.
For each program we considered, the size of the {\it{}Used Area} bar represents
the total memory space occupied by used areas of blocks which are filled gradually.
The {\it{}Supply Area} bar represents the total memory space occupied by supply
areas of blocks which are filled gradually.
Then, the {\it{}No Gradual Filling} bar corresponds to the memory used by arrays which 
are not filled progressively.
Figure \ref{gimp-gcc-etc-1} shows that gradual filling is a quite common phenomenon
and that many software could benefit from the GC optimization presented before.
The best result is obtained by {\tt{}gcc} with 44.4\% of supply area (i.e. ignored by the
GC), which is a similar result to what we found for SmartEiffel (42\%).
Similarly, {\tt{}evince} (29.4\%), {\tt{}unzip} (31.5\%) and {\tt{}gimp} (39.3\%) are 
also good candidates to benefit from this optimization.
Also in Figure \ref{gimp-gcc-etc-1}, we see that {\tt{}vlc} might not benefit from our
optimization because 96.5\% of arrays memory is not filled gradually. 
Finally, for {\tt{}dvips}, even if arrays which are filled gradually represent 96.3\%, our
optimization would be useless because there is no supply area.
Note that measurements were made just before the programs exit.
Knowing that a GC cycle may occur at any time in the life of a program, the situation can be
completely different (larger supply areas or arrays which are not yet allocated).

When an array is filled gradually, it is not necessary or useful to initialize
memory when creating a new array.
Thus, in C, it is preferable to use the {\tt{}malloc} function rather than 
the {\tt{}calloc} function which is more expensive (i.e. in addition to what
is done by the {\tt{}malloc} function, the {\tt{}calloc} function clears the
allocated memory).
We performed a re-run of tests of figure \ref{gimp-gcc-etc-1} under the same conditions 
in order to inspect how arrays are allocated ({\tt{}malloc} or {\tt{}calloc}).
Figure \ref{gimp-gcc-etc-2} shows the results in number of allocated blocks.
The {\it{}malloc} bar represents the number of blocks which have been filled
gradually and, allocated with {\tt{}malloc}.
The {\it{}calloc} bar represents the number of blocks which have been filled
gradually and, allocated with {\tt{}calloc}.
Then, the {\it{}No Gradual Filling} bar is for all other arrays of pointers.
The results in figure \ref{gimp-gcc-etc-2} show that the vast majority 
of arrays that are filled gradually are, which is consistent, 
allocated with {\tt{}malloc}.
Note however that for {\tt{}evince}, 11.3\% of arrays are filled gradually and are 
built with {\tt{}calloc}.
Similarly, {\tt{}emacs} (8.9\%) and {\tt{}gcc} (7.7\%) have a significant number of
arrays that might be allocated more efficiently using {\tt{}malloc} instead of
{\tt{}calloc}.
One could well imagine that the memory tracking tools (e.g. Valgrind), which are already
looking for uninitialized memory usages, could be enhanced with the detection of progressive
filling up, in order to suggest the programmer to advantageously replace some calls to 
{\tt{}calloc} by calls to {\tt{}malloc}.

\subsection{Memory Behavior Analysis of Some Other Java Programs}\label{sub-bench-dacapo}
\begin{figure}
\begin{center}
\includegraphics[scale=1.0]{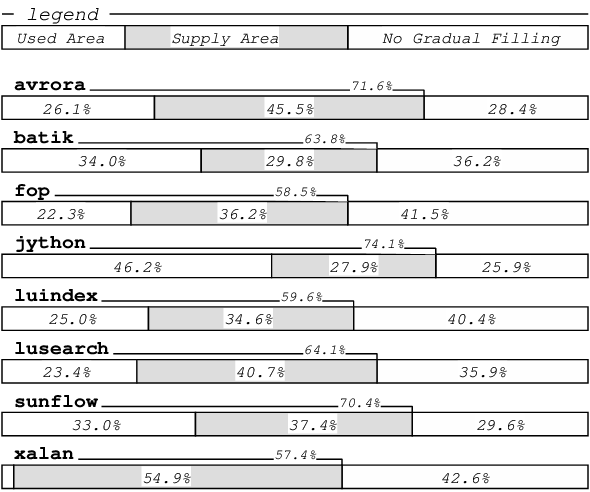}
\caption{Java arrays benchmark. Modified Jikes {\tt{}RVM 3.1.3} and DaCapo {\tt{}9.12-bach}. Detection of memory blocks which are filled gradually. Relative memory used for all arrays of references.}
\label{dacapo}
\end{center}
\end{figure}
To complete our tests, we measured the behavior of Java programs 
as regards the handling of arrays of references.
As before, the aim is to know if arrays are filled gradually or not.
We modified the code of the Java Virtual Machine (Jikes RVM 3.1.3) in order to trace 
all write accesses inside arrays of references.
Thus, we have equipped the {\tt{}aastore} instruction of the virtual machine which is 
the necessary step of all write operations inside arrays of references.
As in previous tests, we only detected the gradual left-to-right filling up of arrays, 
throughout the program execution.
We used the DaCapo benchmarks version 9.12-bach \cite{dacapo2006}.
Figure \ref{dacapo} shows the results in terms of memory space.
The {\it{}Used Area} (resp. {\it{}Supply Area}) bar represents the total memory space occupied by used
areas (resp. supply areas) of blocks which are filled gradually.
The {\it{}No Gradual Filling} bar corresponds to the memory used by arrays which 
are not filled progressively.
The results presented in Figure \ref{dacapo} clearly shows that a very large proportion of 
arrays are filled gradually and that the supply area is very high (from 27.9\% for {\tt{}jython} to 
54.9\% {\tt{}xalan}).
Also note that, as before, the measurement was made on completion (i.e. just before the final exit).

\section{Discussion and Related Work}\label{sec-discuss-related}
\subsection{Related Work for Type Analysis Inside Arrays}
Most of the work done about arrays concerns array bound checks
\cite{cousot1976a,gupta1990a,chin1995a}.
The automatic analysis of properties of array content was considered only 
recently.
Works in \cite{bradley2006a,iosif2008a} study decidable logics to express
properties of array content.
If we restrict ourselves to automatic analysis, an important track initiated
by \cite{flanagan2002a} concerns verification of programs with arrays using
predicate abstraction \cite{lahiri2003a,lahiri2004a}, possibly improved with 
counter-example guided refinement \cite{beyer2007a} and 
Graig interpolants \cite{jhala2007a}.
All these approaches make use of the property to be proved, while 
\cite{halbwachs2008a} aims at discovering properties.

Indeed, the work presented in \cite{halbwachs2008a} concerns properties for 
one-dimensional arrays in non-trivial programs, which manipulate arrays only
by sequential traversal when the array content is restricted to scalar 
(i.e. primitive) values.
Discovered properties in \cite{halbwachs2008a} are:
array maximum value detection or array copy detection, or even
correct insertion of some value inside a sorted array.
For object-oriented languages, it is important to address polymorphism inside
arrays as well as {\sc{}null} prediction inside arrays.
Indeed, it is quite common to have multiple kinds of objects mixed together
with possible {\sc{}null} values inside the same array.

Concerning automatic program analysis methods based on abstract interpretation,
a common approach presented in \cite{blanchet2003a}, for safety critical
software, consists in summarizing an array with one auxiliary variable, managed
to satisfy the disjunction of properties of all cells of the array.
We adapted the {\it{}array smashing} method of \cite{blanchet2003a} for type
flow analysis and, thanks to our gradual filling strategy, we solved its
initialization problem pointed out in \cite{halbwachs2008a}.

While traditional implementations of arrays use contiguous storage, 
\cite{siebert2000a,bacon2003a,chen2003a,pizlo2010a, sartor2010a} are splitting the array 
in memory by using a {\it{}discontinuous storage}.
To reduce fragmentation, \cite{siebert2000a} organises array memory in trees.
To avoid tree traversal, \cite{bacon2003a,pizlo2010a} use a single level of indirection
to fixed-size {\it{}arraylets}.
\cite{chen2003a} contemporaneously invented arraylets
to aggressively compress arrays during collection and decompress
on demand for memory-constrained embedded systems.
To remove most indirections, Z-rays of \cite{sartor2010a} inlines the first 
$N$ array bytes into the array spine.
For arrays of small sizes, the most numerous category, Z-ray avoids indirections and 
reaches the performance we have for the contiguous layout.
Somehow, with Z-rays, the right side of arrays (i.e. parts with indirection) looks like
our supply area.
However, Z-rays storage is inherent in two properties of the target language 
(e.g. Java): zero-initialization of all arrays to take advantage of the zero-compression and
the abstraction of the physical representation of an array (i.e. no pointer manipulation for
array traversal).

\subsection{About the Initialization of Arrays and Variables}
An uninitialized variable is a well known origin of bugs: unpredictable and/or 
randomly initialized integers, invalid pointers to wrong memory locations, etc. 
Furthermore, an uninitialized variable adds uncertainty which makes
type flow analysis impossible or at least useless.
As a consequence, to achieve type flow analysis, the language 
must, either force the programmer to initialize all variables, or provide 
default values for uninitialized variables.
The C\# and Java languages provide default values for all instance/class
variables and force the initialization of all local variables.
For the vast majority of new programming languages, uninitialized variables 
are avoided, either with languages default rules, or with compile time data
flow checks.
For less recent languages (e.g. Fortran, Basic or C), it is also possible to
force the initialization thanks to external extra tools, thus allowing type flow analysis.

Notice that for a language like Java, without default value for local 
variables, the {\sc{}null} situation implies that an explicit assignment 
with {\sc{}null} exists or, by transitivity, that another assigned 
expression could hold that value.

In a language like Eiffel, the default is {\sc{}null} for all kinds of
reference variables, i.e. local variables or instance variables which are
not of a scalar type.
As a drawback, this uniform language decision forces the use of {\it{}data 
flow analysis} to predict that a certain variable cannot hold {\sc{}null} in 
its type set.
Here, it is necessary to take into account the order of statements, just after
the local variable declaration, or inside all creation procedures for an
instance variable, in order to guarantee that the default value is always
overwritten before the variable is read.
Insofar as this language decision can be considered a facility from the
programmer's point of view, it makes the work of implementing type 
analysis harder.
We believe that explicit initialization of all variables should be the rule
as is the case in our initialization strategy of arrays.
At least, this makes it easy to perform type flow analysis.

\section{Conclusion}\label{sec-conclusion}

We have shown how it is possible to optimize garbage collection of arrays by
ignoring the supply area, usually present in flexible/resizable arrays.
We have considered several kinds of resizable arrays: left-most zero indexed,
user defined indexing, circular arrays and hash maps.
Each kind of array is described as an abstract data type also including
a customized marking procedure for the garbage collector.


We have presented a simple technique to analyze the type of elements that are stored in arrays.
That analysis does not take into account either the order of elements in the array or the index
variation.
The whole array, composed of many cells, is considered as a single polymorphic variable.
We have shown that the gradual filling up of the array, cell by cell, facilitates
detection of the {\sc{}null} value inside the array.

We measured the impact of our array manipulation strategy during the bootstrap of
the SmartEiffel compiler. 
On that large piece of code, our technique avoids scanning 42\% of the memory used for 
arrays of references (see section \ref{sub-bench-eiffel}).

Thanks to our technique of type flow analysis inside all kinds of arrays, 
our Lisaac compiler 
is able to carry out a true global analysis.
Roughly, we measured that 98\% of the variables do benefit from array
type flow information.
We also measured that 56\% of arrays never hold {\sc{}null} values.
Furthermore, the level of polymorphism in arrays is significantly reduced
(see section \ref{sub-bench-lisaac}).

We also measured on some other large well-known applications, written in {\sc{}C} 
or \cpp{}, that the ratio of arrays using gradual filling is often similar to the
one encountered in SmartEiffel or Lisaac (see section \ref{sub-bench-other}).

\ack{The authors wish to thank Jean-Pierre Camal and Vasilica Le Floch for proofreading.}

\bibliographystyle{abbrvnat}


\end{document}